%% file: main.tex
\documentclass[conference]{IEEEtran}
\IEEEoverridecommandlockouts

\usepackage[nolist,nohyperlinks]{acronym}
\usepackage{graphicx}
\usepackage[utf8]{inputenc}
\usepackage{listings}
\usepackage{amsmath}
\usepackage{amssymb}
\usepackage[margin=1.0in]{geometry}
\usepackage[colorlinks=true,citecolor=black,linkcolor=black]{hyperref}
\DeclareMathOperator*{\argmax}{arg\,max}

\graphicspath{ {./images/} }

\newlength{\bibitemsep}
\newlength{\bibparskip}
\let\oldthebibliography\thebibliography
\renewcommand\thebibliography[1]{%
\oldthebibliography{#1}%
\setlength{\parskip}{\bibitemsep}%
\setlength{\itemsep}{\bibparskip}%
}

\mathcode`;=\numexpr\mathcode`;-"6000

\begin{acronym}
    \input{sapacronyms.txt}
\end{acronym}

\begin{document}
\title{The Neural-SRP method \\ for positional sound source localization
\thanks{
The research leading to these results has received funding from the European Union's Horizon 2020 research and innovation programme under the Marie Skłodowska-Curie grant agreement No. 956962 and from the European Research Council under the European Union's Horizon 2020 research and innovation program / ERC Consolidator Grant: SONORA (no. 773268). This paper reflects only the authors' views and the Union is not liable for any use that may be made of the contained information.
}}
\author{\IEEEauthorblockN{Eric Grinstein\IEEEauthorrefmark{1}, Toon van Waterschoot\IEEEauthorrefmark{2}, Mike Brookes\IEEEauthorrefmark{1} and Patrick A. Naylor\IEEEauthorrefmark{1}}
\IEEEauthorblockA{\IEEEauthorrefmark{1}Department of Electrical and Electronic Engineering, Imperial College London, U.K.}
\IEEEauthorblockA{\IEEEauthorrefmark{2}Department of Electrical Engineering (ESAT), KU Leuven, Belgium\\
Email: e.grinstein@imperial.ac.uk}
}

\maketitle
\thispagestyle{plain}
\pagestyle{plain}


\maketitle

\input{sections/1_introduction}

\input{sections/2_literature_review}

\input{sections/3_method}

\input{sections/4_experimentation}

\input{sections/5_results_and_discussion_and_conclusion}

\bibliography{sapstrings,Grinstein2023b}{}
\bibliographystyle{plain} 

\end{document}

%% file: sapacronyms.txt
\acro{3GPP}{3rd Generation Partnership Project}
\acro{a-SLAM}[aSLAM]{Acoustic \aclu{SLAM}}
\acro{aSLAM}{Acoustic \aclu{SLAM}}
\acro{A-SNR}{A-weighted \acs{SNR}}
\acro{AAC}{Advanced Audio Coding\acroextra{. A lossy codec used for digital audio.}}
\acro{AAD}{Auditory Attention Detection}
\acro{AAS}{American Auditory Soc.}
\acro{AASP}{Audio and Acoustic Signal Processing}
\acro{ABC}{Analytical with or without Bias Compensation}
\acro{ABR}{Auditory-Brainstem Response}
\acro{ACAWD}{Archivable Core Actual-Word Database}
\acro{ACB}{Adaptive Codebook}
\acro{ACC}{Accuracy}
\acro{ACE}{Acoustic Characterization of Environments\acroextra{. A noisy reverberant speech corpus and IEEE challenge run by the SAP group at Imperial College}}
\acro{ACELP}{Algebraic Code-Excited Linear Prediction}
\acro{ACF}{Autocorrelation Function}
\acro{ACK}{Acknowledgement}
\acro{ACL}{Access Control List}
\acro{ACR}{Absolute Category Rating}
\acro{AD}{Audio Diarization}
\acro{ADC}{Analogue-to-Digital Converter}
\acro{ADM}{Adaptive Differential Microphone}
\acro{ADPCM}{Adaptive Differential Pulse Code Modulation}
\acro{ADSL}{Asymmetric Digital Subscriber Line}
\acro{AE}{Almost Everywhere}
\acro{AES}{Audio Engineering Society}
\acro{AES2}[AES]{Advanced Encryption Standard}
\acro{AGC}{Automatic Gain Control}
\acro{AH}{Amplitude Histogram}
\acro{AI}{Articulation Index}
\acro{AI2}[AI]{Artificial Intelligence}
\acro{AI3}[AI]{Audio Inpainting}
\acro{AIC}{Akaike Information Criterion}
\acro{AIFF}{Audio Interchange File Format}
\acro{AIR}{Acoustic Impulse Response}
\acro{AIR2}[AIR]{Aachen Impulse Response}
\acro{AIRD}{Aachen Impulse Response Database}
\acro{ALC}{Automatic Level Control}
\acro{ALCons}{Articulation Loss of Consonants}
\acro{AM}{Amplitude Modulation}
\acro{AMDF}{Average Magnitude Difference Function\acroextra{. A function with similar properties to the cross- or autocorrelation but that requires no multiplication to evaluate.}}
\acro{AMR}{Adaptive Multi-Rate}
\acro{AMR-NB}{Adaptive Multi-Rate Narrow Band}
\acro{AMR-WB}{Adaptive Multi-Rate Wide Band}
\acro{AMS}{Amplitude Modulation Spectrogram}
\acro{ANC}{Adaptive Noise Canceller}
\acro{ANS}{Autocorrelation-Based Noise Subtraction}
\acro{ANSI}{American National Standards Institute}
\acro{ANU}{Australian National University}
\acro{APLAWD}{Archivable Priority List Actual-Word Database}
\acro{AR}{Autoregressive}
\acro{ARD}{Arbeitsgemeinschaft der \"{o}ffentlich-rechtlichen Rundfunkanstalten der Bundesrepublik Deutschland\acroextra{. `Consortium (``Working group'') of the public-law broadcasting institutions of the Federal Republic of Germany'}}
\acro{ARMA}{Autoregressive Moving Average}
\acro{AS}{Audio Segmentation}
\acro{AS2}[AS]{Almost Surely}
\acro{ASA}{Acoustic Scene Analysis}
\acro{ASIO}{Audio Stream Input/Output\acroextra{. A computer soundcard protocol with low latency developed by Steinberg.}}
\acro{ASK}{Amplitude Shift Keying}
\acro{ASLP}{Audio, Speech, and Language Processing}
\acro{ASM}{Acoustic Scene Mapping}
\acro{ASR}{Automatic Speech Recognition}
\acro{ASS}{Approximate Spectrum Substitution}
\acro{AST}{Acoustic Source Tracking}
\acro{AST2}[AST]{Asymmetric Sampling in Time}
\acro{ATF}{Acoustic Transfer Function\acroextra{. The Fourier Transform of the \acs{RIR}.}}
\acro{ATLM}{Acoustic Tokenization and Language Modelling}
\acro{ATR}{Advanced Telecommunications Research Institute International\acroextra{, Kyoto, Japan}}
\acro{AUC}{Area under the curve}
\acro{AURORA}{Aurora Experimental Framework for the Evaluation of the Performance of Speech Recognition Systems under Noisy Conditions}
\acro{AUV}{Autonomous Underwater Vehicle}
\acro{AV}{Audio-Visual}
\acro{AWGN}{Additive White Gaussian Noise}
\acro{AWS}{Approximate Waveform Substitution}
\acro{BASIE}{Bayesian Adaptive Speech Intelligibility Estimation}
\acro{BCE}{Blind Channel Estimation}
\acro{BCL}{Bekesy Comfortable Loudness}
\acro{BCR}{Block-Coordinate Relaxation}
\acro{BEM}{Boundary Element Method}
\acro{BER}{Bit Error Rate}
\acro{BIBO}{Bounded-Input Bounded-Output}
\acro{BIC}{Bayesian Information Criterion}
\acro{Bk}{Berksons\acroextra{. A unit for measuring intelligibility.}}
\acro{BK}[B\&K]{Br{\"u}el and Kj{\ae}r}
\acro{BLSTM}{Bidirectional \acs{LSTM}}
\acro{BM}{Blocking Matrix}
\acro{BO-SCPHD}{Bearing-only \acs{SC-PHD}}
\acro{BO-SLAM}{Bearing-only \acs{SLAM}}
\acro{BOT}{Bearing-only tracking}
\acro{BP}{Basis Pursuit}
\acro{BPCC}{Basis Pursuit with Clipping Constraints}
\acro{BPDN}{Basis Pursuit Denoising}
\acro{BPM}{Beats Per Minute}
\acro{BPSK}{Binary Phase Shift Keying}
\acro{BR}{Barrodale and Roberts' (algorithm)}
\acro{BRI}{Basic Rate Index}
\acro{BSD}{Bark Spectral Distortion}
\acro{BSI}{Blind System Identification}
\acro{BSIM}{Binaural Speech Intelligbility Model}
\acro{BSS}{Blind Source Separation}
\acro{BSTOI}{Binaural \acs{STOI}}
\acro{BW}{Bandwidth}
\acro{BZ}{Back-to-Zero}
\acro{C4DM}{Centre for Digital Music}
\acro{C50}[$C_\textrm{50}$]{Clarity Index}
\acro{C-GFB}{Combination Gas-fired Boiler}
\acro{CART}{Classification and Regression Tree}
\acro{CASA}{Computational Auditory Scene Analysis}
\acro{CBR}{Constant Bit Rate}
\acro{CCC}{Cross-Correlation Coefficient}
\acro{CCCC}{DARPA CSR Corpus Coordinating Committee}
\acro{CCD}{Charge-Coupled Device}
\acro{CCI}{Call Clarity Index}
\acro{CCITT}{Consultative Committee for International Telephony and Telegraphy}
\acro{CCM}{Contralateral Competing Message}
\acro{CCR}{Comparison Category Rating}
\acro{CDB}{Constant Directivity Beamformer}
\acro{CDMA}{Code Division Multiple Access}
\acro{CELP}{Code-excited Linear Prediction}
\acro{CHIEF}{Combined Helmholtz Integral Equation Formulation}
\acro{CIT}{Constrained Initial Taps}
\acro{CL}{Clipping Level}
\acro{CLEAR}{Centre for Law Enforcement Audio Research}
\acro{CLID}{Cluster Identification Test}
\acro{CLT}{Central Limit Theorem}
\acro{CMA}{Constant Modulus Algorithm}
\acro{CMASI}{Coherence Modulated Acoustic Speckle Interferometry}
\acro{CMB}{Cosmic Microwave Background}
\acro{CNC}{Consonant-Nucleus-Consonant}
\acro{CNG}{Comfort Noise Generation}
\acro{CNN}{Convolutional Neural Network}
\acrodefplural{CNN}[CNNs]{Convolutional Neural Networks}
\acro{CODEC}{Coder-Decoder}
\acro{CPE}{Customer Premises Equipment}
\acro{CPHD}{Cardinalized \acs{PHD}}
\acro{CRACD}{Codec-robust Automatic Clipping Detector}
\acro{CRC}{Cyclic Redundancy Check}
\acro{CRNN}{Convolutional Recurrent Neural Network}
\acrodefplural{CRNN}[CRNNs]{Convolutional Recurrent Neural Networks}
\acro{CS}{Channel Shortening}
\acro{CS2}[CS]{Compressive Sensing}
\acro{CSP}{Communications and Signal Processing}
\acro{CSR-WSJ}{Continuous Speech Recognition Wall Street Journal Phase 1\acroextra{ database}}
\acro{CST}{Connected Speech Test\acroextra{ speech corpus}}
\acro{CT}{Conversation Test}
\acro{CTTN}{Comparative Tolerance to Noise}
\acro{CV}{Coefficient of Variation}
\acro{CVNN}{Complex-Valued Neural Networks}
\acrodefplural{CV}{Coefficients of Variation}
\acro{CV2}[CV]{Constant Velocity}
\acro{CVC}{Consonant-Vowel-Consonant}
\acro{CW}{Continuous Wave}
\acro{CWM}{Centre-Weighted Median}
\acro{CWMY}{Centre-Weighted Myriad}
\acro{CWT}{Continuous Wavelet Transform}
\acro{DAC}{Digital-to-Analogue Converter}
\acro{DAM}{Diagnostic Acceptability Measure}
\acro{DAQ}{Data Acquisition}
\acro{DARPA}{Defense Advanced Research Projects Agency\acroextra{ of the United States Dept. of Defense}}
\acro{DARPA-RMD}{\acs{DARPA} 1000-Word Resource Management Database\acroextra{ for Continuous Speech Recognition}}
\acro{DAW}{Digital Audio Workstation}
\acro{dB}{Decibel}
\acro{dBFS}{\acs{dB} Full Scale}
\acro{DBN}{Deep Belief Network}
\acro{DBSTOI}{Deterministic \acs{BSTOI}}
\acro{DC}{Direct Current}
\acro{DCME}{Digital Circuit Multiplexing Equipment}
\acro{DCR}{Degradation Category Rating}
\acro{DCT}{Discrete Cosine Transform}
\acro{DDR}{Direct-to-diffuse ratio}
\acro{DDR3}{Double Data Rate Type Three}
\acro{DECT}{Digital European Cordless Telecommunication}
\acro{DeLILAH}{Detection of Clipping using Least Squares Residuals and Iterated Logarithm Amplitude Histogram}
\acro{DENBE}{\acs{DRR} Estimation using a Null-Steered Beamformer}
\acro{DET}{Detection Error Trade-off}
\acro{Dev}{Development\acroextra{ dataset of the \acs{ACE} Challenge}}
\acro{DFT}{Discrete Fourier Transform}
\acro{DI}{Directivity Index}
\acro{DirectX}{\acroextra{A programming interface developed by Microsoft for handling tasks related to multimedia.}}
\acro{DIRHA}{Distant-speech Interaction for Robust Home Applications\acroextra{ multi-microphone multi-language acoustic speech corpus}}
\acro{DMA}{Distributed Microphone Array}
\acro{DMT}{Discrete Multi-Tone}
\acro{DMV}{Dynamically Managed Voice\acroextra{ system}}
\acro{DNN}{Deep Neural Network}
\acro{DNR}{Dynamic Noise Reduction}
\acro{DoA}{Direction-of-Arrival}
\acrodefplural{DoA}[DoAs]{Directions-of-Arrival}
\acro{DOA}{Direction-of-Arrival}
\acrodefplural{DOA}[DOAs]{Directions-of-Arrival}
\acro{DP}{Dynamic Programming}
\acro{DPCM}{Differential Pulse Code Modulation}
\acro{DPD}{Direct-Path Dominance}
\acro{DPD-MUSIC}{Direct-Path Dominance Multiple Signal Classification}
\acro{DR}{Douglas-Rachford}
\acro{DR2}{Dynamic Range}
\acro{DRM}{Diagnostic Rhyme Test}
\acro{DRR}{Direct-to-Reverberant Ratio}
\acro{DRNN}{Deep Recurrent Neural Net}
\acro{DRT}{Diagnostic Rhyme Test}
\acro{DSB}{Delay-and-Sum Beamformer}
\acro{DSOBM}{Deterministic \acs{SOBM}}
\acro{DSP}{Digital Signal Processing}
\acro{DSPS}{Double Sides Periodic Substitution}
\acro{DSR}{Distributed Speech Recognition}
\acro{DSWS}{Double Sides Waveform Substitution}
\acro{DTMF}{Dual Tone Multi-Frequency}
\acro{DTX}{Discontinued Transmission}
\acro{DWT}{Discrete Wavelet Transform}
\acro{EARS}{Embodied Audition for RobotS}
\acro{EBF}{Eigen-beamformer}
\acro{EBU}{European Broadcasting Union}
\acro{EC}{Echo Canceller}
\acro{EDC}{Energy Decay Curve}
\acro{EDR}{Energy Decay Relief}
\acro{EDF}{Energy Decay Function}
\acro{EEG}{Electroencephalography}
\acro{EER}{Equal Error Rate}
\acro{EFICA}{Efficient Fast Independent Component Analysis}
\acro{EIR}{Equalized Impulse Response}
\acro{EKF}{Extended Kalman Filter}
\acro{EKF-SLAM}[EKF-SLAM]{\acs{EKF} \acs{SLAM}}
\acro{EKF-SLAM2}[EKF-SLAM]{Extended Kalman Filter \acs{SLAM}}
\acro{EL}{Echo Loss}
\acro{ELF}{Extremely Low Frequency}
\acro{ELRA}{European Languages Research Association}
\acro{EM}{Estimation-Maximization\acroextra{. An iterative technique to solve certain optimization problems.}}
\acro{EMIB}{Eigenmike Microphone Interface Box}
\acro{ENF}{Electrical Network Frequency}
\acro{EPSRC}{Engineering and Physical Sciences Research Council}
\acro{EQ}{Equalisation}
\acro{ERB}{Equivalent Rectangular Bandwidth}
\acro{ERP}{Ear Reference Point (cf. ITU-T Rec. P.64 1999)}
\acro{ESA}{Early Stage Assessment}
\acro{ESM}{Equivalent Source Method}
\acro{ESPRIT}{Estimation of Signal Parameters via Rotational Invariance Techniques}
\acro{ETAN}{Equivalent Tolerance to Additional Noise} 
\acro{ETSI}{European Telecommunications Standards Institute}
\acro{EURASIP}{European Association for Signal Processing}
\acro{EUSIPCO}{European Signal Processing Conference}
\acro{Eval}{Evaluation\acroextra{ dataset of the \acs{ACE} Challenge}}
\acro{F1}{F1 Score}
\acro{FAR}{False Alarm Rate}
\acro{FastSLAM}[FastSLAM]{FActored Solution To Simultaneous Localization and Mapping}
\acro{FastSLAM2}[FastSLAM]{FActored Solution To \acs{SLAM}} 
\acro{FAU}{Friedrich-Alexander-Universit{\"a}t}
\acro{FB}{Forward-Backward}
\acro{FB2}[FB]{Fullband}
\acro{FBF}{Fixed Beamformer}
\acro{FBSS}{Forward-Backward Spatial Smoothing}
\acro{FCC}{Federal Communications Commission}
\acro{FDM}{Frequency Division Multiplexing}
\acro{FDR}{False Discovery Rate}
\acro{FDR2}[FDR]{Free-Decay Region}
\acro{FEC}{Forward Error Correction}
\acro{FEM}{Finite Element Method}
\acro{FFI}{Norwegian Defence Research Establishment}
\acro{FFT}{Fast Fourier Transform}
\acroindefinite{FFT}{an}{a}
\acro{FIFO}{First-In First-Out}
\acro{FIR}{Finite Impulse Response\acroextra{. A filter whose output is a weighted sum of past input values and whose system function contains only zeros and no poles.}}
\acro{FISM}{Fast Image Source Method}
\acro{FISST}{Finite Set STatistics}
\acro{FLOM}{Fractional Lower-Order Moments}
\acro{FLOS}{Fractional Lower-Order Statistics}
\acro{FM}{Frequency Modulation}
\acro{FMM}{Fast Multipole Method}
\acro{FN}{False Negative}
\acro{FN2}{Nth Formant}
\acro{FNR}{False Negative Rate}
\acro{FORTRAN}{The IBM Mathematical Formula Translating System}
\acro{FoV}{Field of View}
\acro{FP}{False Positive}
\acro{FPR}{False Positive Rate}
\acro{FPS}{Frames Per Second}
\acro{FRI}{Finite Rate of Innovation}
\acro{FSB}{Filter-and-Sum Beamformer}
\acro{FSK}{Frequency Shift Keying}
\acro{FT}{Flat-Top}
\acro{FWER}{Familywise Error Rate}
\acro{FWSSNR}{Frequency-Weighted Segmental \acs{SNR}}
\acro{FWSSRR}{Frequency-Weighted \acs{SSRR}}
\acro{G.711}{\acs{PCM} of Voice Frequencies}
\acro{GARCH}{Generalized Auto-regressive Conditional Heteroscedasticity}
\acro{GBW}{Gain Bandwidth Product}
\acro{GCC}{Generalized Cross-Correlation}
\acro{GCC-PHAT}{Generalized Cross-Correlation with Phase Transform\acroextra{ method of estimating \acs{TDoA}}}
\acro{GCF}{Global Coherence Field}
\acro{GGD}{Generalized Gaussian Distribution}
\acro{GGD2}[G$\Gamma$D]{Generalised Gamma Distribution}
\acro{GM}{Gaussian Mixture}
\acro{GM-PHD}{Gaussian Mixture \acs{PHD}}
\acro{GMCA}{Generalized Morphological Component Analysis}
\acro{GMM}{Gaussian Mixture Model\acroextra{. An approximation to an arbitrary probability density function that consists of a weighted sum of Gaussian distributions}}
\acro{GNN}{Graph Neural Network}
\acro{GNSS}{Global Navigation Satellite System}
\acro{GPRS}{General Packet Radio Services}
\acro{GPS}{Global Positioning System}
\acro{GRU}{Gated Recurring Unit}
\acro{GSC}{Generalized Sidelobe Canceller}
\acro{GSM}{Global System for Mobile Communications}
\acro{GSM-EFR}{\acs{GSM} Enhanced Full Rate Codec}
\acro{GSM-FR}{\acs{GSM} Full Rate Codec}
\acro{GSM-HR}{\acs{GSM} Half Rate Codec}
\acro{GUI}{Graphical User Interface}
\acro{HAAC}{High Amplitude Audio Capture}
\acro{HATS}{Head and Torso Simulator}
\acro{HERB}{Harmonicity-based dEReverBeration}
\acro{HFT}{Hands-Free Terminal}
\acro{HI}{Hearing-Impaired}
\acro{HIE}{Helmholtz Integral Equation}
\acro{HISAS}{High resolution Interferometric \ac{SAS}}
\acro{HINT}{Hearing-in-Noise Test}
\acro{HLT}{Human Language Technology}
\acro{HMM}{Hidden Markov Model}
\acro{HOS}{Higher-Order Statistics}
\acro{HPF}{High-Pass Filter}
\acro{HR}{Half Rate (\acs{GSM} Codec)}
\acro{HRI}{Human-Robot Interaction}
\acro{HRTF}{Head-related Transfer Function}
\acro{HSA}{Hearing, Speech, Audio\acroextra{ technology group at Fraunhofer IDMT}}
\acro{HSD}{Hybrid Steepest Descent}
\acro{HT}{Hannan-Thomson}
\acro{HTK}{Hidden Markov Model Tool Kit}
\acro{IBM}{Ideal Binary Mask}
\acro{IC}{Interference Canceller}
\acro{ICA}{Independent Component Analysis}
\acro{ICASSP}{Intl. Conf. on Acoustics, Speech and Signal Processing}
\acro{ID}{Identifier}
\acro{IDMT}{Institute for Digital Media Technology}
\acro{iDEN}{Integrated Digital Enhanced Network}
\acro{IEC}{International Electrotechnical Commission}
\acro{IEEE}{Institute of Electrical and Electronics Engineers}
\acro{IET}{Institute of Engineering and Technology}
\acro{IETF}{Internet Engineering Task Force}
\acro{IFFT}{Inverse Fast Fourier Transform}
\acro{IHC}{Inner Hair Cell}
\acro{IHT}{Iterative Hard Thresholding}
\acro{IID}{Independent and Identically Distributed}
\acro{IIR}{Infinite Impulse Response\acroextra{. A filter whose output is a weighted sum of both past input and past output values and whose system function contains both poles and zeros.}}
\acro{IL}{Iterated Logarithm\acroextra{, the logarithm of the logarithm}}
\acro{ILAH}{Iterated Logarithm Amplitude Histogram\acroextra{ clipping detection method}}
\acro{ILD}{Interaural Level Difference}
\acro{IMCRA}{Improved Minima Controlled Recursive Averaging\acroextra{. A technique for blindly estimating the spectrum of additive noise in a signal.}}
\acro{IMD}{Inter-Modulation Distortion}
\acro{IMSI}{International Mobile Subscriber Identity}
\acro{IMU}{Inertial Measurement Unit}
\acro{INMD}{In-service Non-intrusive Measurement Device}
\acro{INTERSPEECH}{Annual Conference of the \acs{ISCA}}
\acro{IO}{Infinitely Often}
\acro{IP}{Internet Protocol}
\acro{IPA}{International Phonetic Association}
\acro{IPD}{Interaural Phase Difference}
\acro{IRM}{Ideal Ratio Mask}
\acro{IRS}{Inverse repeated Sequence\acroextra{. A pseudo random sequence used for impulse response measurement.}}
\acro{IRS2}[IRS]{Intermediate Reference System}
\acro{IS}{Importance Sampling}
\acro{ISCA}{International Speech Communication Association}
\acro{ISDN}{Integrated Services Digital Network}
\acro{ISFT}{Inverse \acl{SFT}}
\acro{ISO}{Intl. Organization for Standardization}
\acro{ISP}{Intensity Spectral Profile}
\acro{IST}{Iterative Soft Thresholding}
\acro{ISTFT}{Inverse Short Time Fourier Transform}
\acro{ISVR}{Institute of Sound and Vibration Research\acroextra{, Southampton University, UK}}
\acro{ITD}{Interaural Time Difference}
\acro{ITF}{Interaural Transfer Function}
\acro{ITU}{International Telecommunication Union}
\acro{ITUR}[ITU-R]{International Telecommunication Union\acroextra{ Radiocommunication Sector}}
\acro{ITUT}[ITU-T]{International Telecommunication Union\acroextra{ Telecommunication Standardisation Sector}}
\acro{IUWT}{Isotropic Undecimated Wavelet (Starlet) Transform}
\acro{IWAENC}{Intl. Workshop on Acoustic Signal Enhancement}
\acro{IWAENC_PRE_2012}[IWAENC]{Intl. Workshop Acoustic Echo and Noise Control}
\acro{IWASE}{Intl. Workshop on Acoustic Signal Enhancement}
\acro{JADE}{Joint Approximate Diagonalization of Eigen-Matrices}
\acro{JPDA}{Joint Probabilistic Data Association}
\acro{JPEG}{Joint Photographic Experts Group}
\acro{KDE}{Kernel Density Estimate}
\acro{KF}{Kalman Filter}
\acro{KFD}{Kernel Fisher Discriminant\acroextra{ Analysis}}
\acro{KL}{Karhunen-Lo{\'{e}}ve}
\acro{KLT}{Karhunen-Lo{\'{e}}ve Transform}
\acro{KST}{Kolmogorov-Smirnov Test}
\acro{LAD}{Least Absolute Deviation}
\acro{LAN}{Local Area Network}
\acro{LARS}{Least Angle Regression}
\acro{LAT}[L$_{\textrm{AT}}$]{Equivalent Continuous Sound Level\acroextra{. Also called Leq}}
\acro{LBR}{Low Bitrate Redundancy}
\acro{LC}{Local Criterion}
\acro{LCMP}{Linearly Constrained Minimum Power}
\acro{LCMV}{Linearly Constrained Minimum Variance}
\acro{LCWM}{Linear Combination of Weighted Medians}
\acro{LDC}{Linguistic Data Consortium}
\acro{LEM}{Loudspeaker-Enclosure-Microphone System}
\acro{Leq}[L$_{\textrm{eq}}$]{Equivalent Continuous Sound Level\acroextra{. Also called LAT}}
\acro{LF}{Liljencrants-Fant\acroextra{. The developers of a glottal waveform model}}
\acro{LHS}{Left-Hand Side}
\acro{LID}{Language Identification}
\acro{LILAH}{\acs{LSR2}-\acs{ILAH}\acroextra{ clipping detection method}}
\acro{LIME}{LInear Predictive Multi-input Equalization\acroextra{ algorithm}}
\acro{LiNoPS}{Lightweight Noise Protection System}
\acro{LLN}{Law of Large Numbers}
\acro{LLR}{Log-Likelihood Ratio}
\acro{LLS}{Logarithmic Least Squares}
\acro{LMA}{Least Mean Absolute}
\acro{LMS}{Least Mean Squares\acroextra{ adaptive filter}}
\acro{LNA}{Low Noise Amplifier}
\acro{LoS}{Line-of-Sight}
\acro{LOT}{Listening-Only Test}
\acro{LP}{Linear Parameter}
\acro{LP2}[LP]{Linear Predictive}
\acro{LP3}[LP]{Linear Prediction}
\acro{LPC}{Linear Predictive Coding\acroextra{. An autoregressive model of speech production.}}
\acro{LQO}{Listening Quality Objective}
\acro{LREC}{Conf. on Language Resources and Evaluation}
\acro{LS}{Least Squares}
\acro{LSA}{Log Spectral Amplitude}
\acro{LSB}{Lower Side-Band}
\acro{LSB2}[LSB]{Least Significant Bit}
\acro{LSD}{Log Spectral Distortion}
\acro{LSP}{Line Spectrum Pairs}
\acro{LSR}{Late Stage Review}
\acro{LSR2}[LSR]{Least Squares Residuals\acroextra{ clipping detection method}}
\acro{LSRT}{Least Squares Residuals with Thresholding\acroextra{ clipping detection method}}
\acro{LSTM}{Long Short-Term Memory}
\acro{LTASS}{Long Term Average Speech Spectrum}
\acro{LTI}{Linear Time Invariant}
\acro{LTP}{Long Term Prediction}
\acro{LU}{Loudness Unit}
\acro{LUFS}{Loudness Units Full-Scale}
\acro{MA}{Moving Average}
\acro{MAC}{Multiply Accumulate Operation}
\acro{MAD}{Median Absolute Deviation}
\acro{MAE}{Mean Absolute Error}
\acro{MAP}{Maximum \emph{a posteriori}}
\acro{MARDY}{Multichannel Acoustic Reverberation Database at York}
\acro{MARS}{Multivariate Adaptive Regression Splines}
\acro{MBF}{Matched Filter Beamformer}
\acro{MC}{Monte Carlo}
\acro{MCA}{Morphological Component Analysis}
\acro{MCC}{Matthew's Correlation Coefficient}
\acro{MCEQ}{MultiChannel EQualisation}
\acro{MCS}{Multidimensional Colouration Space}
\acro{MDCT}{Modified Discrete Cosine Transform}
\acro{MDL}{Minimum Description Length}
\acro{MDS}{Multidimensional Scaling}
\acro{Mel}{\acroextra{A non-uniform frequency scale corresponding to perceived frequency. It is approximately linear at low frequencies and logarithmic at high frequencies.}}
\acro{MELP}{Mixed Excitation Linear Prediction}
\acro{MFCC}{Mel-frequency Cepstral Coefficients}
\acro{MFS}{Method of Fundamental Solutions}
\acro{MFSK}{Multi-Frequency Shift Keying}
\acro{MHT}{Multi-Hypotheses Tracking}
\acro{MI}{Mutual Information}
\acro{MIMO}{Multiple-Input-Multiple-Output}
\acro{MINT}{Multiple-input/output INverse Theorem}
\acro{MIRS}{Motorola Integrated Radio System}
\acro{MIT}{Massachusetts Institute of Technology}
\acro{MIT-LCS}{Massacchusetts Institute of Technology Laboratory for Computer Science}
\acro{ML}{Maximum Likelihood}
\acro{MLD}{Masking Level Difference}
\acro{MLE}{Maximum Likelihood Estimation}
\acro{ML-TDoA}{Maximum Likelihood Time Difference of Arrival}
\acro{MLMF}{Machine Learning with Multiple Features}
\acro{MLP}{Multi-layer Perceptron}
\acro{MLS}{Maximum Length Sequence\acroextra{ of pseudo random bits.}}
\acro{MMSE}{Minimum Mean Squared Error}
\acro{MMT}{Multiscale Median Transform}
\acro{MNRU}{Modulated Noise Reference Unit}
\acro{MOM}{Mean of Maximum}
\acro{MOS}{Mean Opinion Score}
\acro{MOS-LQO}{Mean Opinion Score - Listening Quality Objective}
\acro{MP}{Matching Pursuit}
\acro{MP3}{\acs{MPEG}-2 Audio Layer III}
\acro{MPEG}{Moving Picture Experts Group}
\acro{MRF}{Markov Random Field}
\acro{MRP}{Mouth Reference Point (cf. ITU-T Rec. P.64 1999)}
\acro{MRT}{Modified Rhyme Test}
\acro{MS}{Minimum Statistics}
\acro{MSB}{Most Significant Bit}
\acro{MSC}{Mean Square Coherence}
\acro{MSE}{Mean Square Error}
\acro{MSN}{Multiple Subscriber Number}
\acro{MSNR}{Maximum \acs{SNR}}
\acro{MTF}{Modulation Transfer Function}
\acro{MTM}{Modified Trimmed Mean}
\acro{MUSCLE}{MeasUred Single-CLustEr}
\acro{MUSHRA}{Multi-stimuli Test with Hidden Reference and Anchor}
\acro{MUSHRAR}{Multi-stimuli Test with Hidden Reference and Anchor for Reverberation}
\acro{MUSIC}{Multiple Signal Classification}
\acro{MVDR}{Minimum Variance Distortionless Response\acroextra{ beamformer}}
\acro{MWF}{Multi-channel Wiener Filter}
\acro{NAH}{Nearfield Acoustic Holography}
\acro{NB}{Narrowband}
\acro{NCM}{Normalized Coherence Metric}
\acro{NH}{Normal-Hearing}
\acro{NIRA}{Non-Intrusive Room Acoustics}
\acro{NISE}{Non-Intrusive \acs{SNR} estimation}
\acro{NISI}{Non-Intrusive Speech Intelligibility Estimation}
\acro{NISQ}{Non-Intrusive Speech Quality Estimation}
\acro{NIST}{National Institute of Standards and Technology}
\acro{NL}{Noise Level}
\acro{NLA}{Non-Linear Approximation}
\acro{NLMS}{Normalized Least Mean Squares\acroextra{ adaptive filter}}
\acro{NMCFLMS}{Normalized Multichannel Frequency Domain Least Mean Square}
\acro{NMF}{Non-negative Matrix Factorization}
\acro{NOISEX-92}{Database to Study the Effect of Additive Noise on Speech Recognition Systems}
\acro{NOIZEUS}{Noisy Speech Corpus for Evaluation of Speech Enhancement Algorithms}
\acro{NOSRMR}{Normalized Overall \acs{SRMR}}
\acro{NOS}{Number of Sources}
\acro{NOSRMR}{Normalized Overall \acs{SRMR}}
\acro{NPM}{Normalized Projection Misalignment}
\acro{NPV}{Negative Predictive Value}
\acro{NR}{Noise Reduction}
\acro{NS}{Noise Suppression}
\acro{NSRMR}{Normalised \acs{SRMR}}
\acro{NSRR}{Normalized Signal-to-Reverberation Ratio}
\acro{NSRMR}{Normalised \acs{SRMR}}
\acro{NSV}{Negative-Side Variance}
\acro{NTP}{Network Time Protocol}
\acro{NV}{Noise-Vocoding}
\acro{OBL}{Octave Band Level}
\acro{ODF}{Overdrive Factor}
\acro{Ofcom}{Office of Communications\acroextra{, the independent regulator and competition authority for the UK communications industries}}
\acro{OFDM}{Orthogonal Frequency Division Multiplexing}
\acro{OHC}{Outer Hair Cell}
\acro{OIM}{Objective Intelligibility Measure}
\acro{OLA}{Overlap-add}
\acro{OM-LSA}{Optimally Modified Log-Spectral Amplitude{ Estimator}}
\acro{OMP}{Orthogonal Matching Pursuit}
\acro{OSI}{Open Systems Interconnection}
\acro{OSPA}{Optimal Subpattern Assignment}
\acro{OSRMR}{Overall \acs{SRMR}}
\acro{PAB-SRMR}{Per acoustic band \acs{SRMR}}
\acro{PALM}{Passive Acoustic Localization and Mapping}
\acro{PAMS}{Perceptual Analysis Measurement System}
\acro{PARCOR}{Partial Correlation Coefficients}
\acro{PB}{Phonetically Balanced}
\acro{PBF}{Positive Boolean Function}
\acro{PCA}{Principal Components Analysis}
\acro{PCM}{Pulse-Code Modulation}
\acro{PDA}{Personal Digital Assistant}
\acro{PDA2}[PDA]{Probabilistic Data Association}
\acro{PDE}{Partial Differential Equation}
\acro{pdf}{Probability Density Function}
\acro{PDF}{Probability Density Function}
\acro{PE}{Parameter Estimation}
\acro{PEASS}{Perceptual Evaluation for Audio Source Separation}
\acro{PEFAC}{Pitch Estimation Filter with Amplitude Compression}
\acro{PESQ}{Perceptual Evaluation of Speech Quality}
\acro{PF}{Psychometric Function}
\acro{pgfl}[p.g.fl.]{Probability Generating Functional}
\acro{PHAT}{Phase Transform}
\acro{PHD}{Probability Hypothesis Density}
\acro{PIP}{Peak-Image Pairing}
\acro{PIV}{Pseudo-Intensity Vector}
\acro{PL}{Pseudo-likelihood}
\acro{PLC}{Packet Loss Concealment}
\acro{PLL}{Phase Locked Loop}
\acro{PLP}{Perceptual Linear Prediction}
\acro{PLR}{Perceived Level of Reverberation}
\acro{PM}{Phase Modulation}
\acro{PMF}{Probability Mass Function}
\acro{PMOS}{Predicted Mean Opinion Score}
\acro{POLQA}{Perceptual Objective Listening Quality Analysis}
\acro{POTS}{Plain Old Telephone Service}
\acro{PPP}{Poisson Point Process}
\acro{PPS}{Pulse-Per-Second}
\acro{PPV}{Positive Predictive Value}
\acro{PReLU}{Parametric Rectified Linear Unit}
\acro{PRLM}{Phoneme Recognition and Language Modelling}
\acro{PRP}{Pair-wise Relative Phase-ratio}
\acro{PSD}{Power Spectral Density}
\acro{PSK}{Phase Shift Keying}
\acro{PSNR}{Peak Signal-to-Noise Ratio}
\acro{PSOLA}{Pitch Synchronous Overlap Add\acroextra{. A method of scaling a signal in time and pitch independently.}}
\acro{PSQM}{Perceptual Speech Quality Measurement}
\acro{PSSL}{Positional \acs{SSL}}
\acro{PSTN}{Public Switched Telephone Network}
\acro{PWD}{Plane-Wave Decomposition}
\acro{PTA}{Pure-Tone Audiology}
\acro{QAM}{Quadrature Amplitude Modulation}
\acro{QILAH}{Quadrisected Iterated Logarithm Amplitude Histogram\acroextra{ clipping detection method}}
\acro{QMF}{Quadrature Mirror Filter}
\acro{QoE}{Quality-of-Experience}
\acro{QoS}{Quality-of-Service}
\acro{QPSK}{Quadrature Phase Shift Keying}
\acro{RADAR}{RAdio Detection And Ranging}
\acro{RASTA}{Relative Spectral}
\acro{RASTA-PLP}{Relative Spectral Perceptual Linear Prediction}
\acro{RASTI}{Room Acoustics Speech Transmission Index\acroextra{ (superseded by STIPA)}}
\acro{RBM}{Restricted Boltzmann Machine}
\acro{RB-PHD}{Rao-Blackwellised \acs{PHD}}
\acro{RBPF}{Rao-Blackwellised Particle Filter}
\acro{RC}{Relative Criterion}
\acro{RDT}[$R_\textrm{DT}$]{Reverberation Decay Tail}
\acro{RDTF}{Relative Direct Transfer Function}
\acro{ReLU}{Rectified Linear Unit}
\acro{RF}{Radio Frequency}
\acro{RFI}{Radio Frequency Interference}
\acro{RFS}{Random Finite Set}
\acro{RHS}{Right-Hand Side}
\acro{RIP}{Restricted Isometry Property}
\acro{RIR}{Room Impulse Response}
\acro{RLS}{Recursive Least Squares\acroextra{ adaptive filter}}
\acro{RLSD}{Relative Log Spectral Distortion}
\acro{RMCLS}{Relaxed Multichannel Least Squares}
\acro{RMCLS-CIT}{Relaxed MultiChannel Least-Squares with Constrained Initial Taps}
\acro{RMS}{Root Mean Square}
\acro{RMSE}{Root Mean Square Error}
\acro{RNN}{Recurrent Neural Net}
\acro{ROC}{Receiver Operating Characteristic}
\acro{ROHC}{Robust Header Compression}
\acro{RP}{Received Pronunciation}
\acro{RPE}{Regular Pulse Excitation}
\acro{RRTF}{Relative Real-Time Factor}
\acro{RS}{Reverberation Suppression}
\acro{RSM}{Reflector Source Method}
\acro{RSS}{Received Signal Strength}
\acro{RSV}{Room Spectral Variance}
\acro{RT}{Reverberation Time}
\acro{RTAN}{Robustness to Additional Noise}
\acro{RTF}{Real-Time Factor}
\acro{RTF2}[RTF]{Room Transfer Function}
\acro{RTF3}[RTF]{Relative Transfer Function}
\acro{RV}{Random Variable}
\acro{RVP}{Recursive Vector Projection}
\acro{RWTH}{Rheinisch-Westf\"{a}lische Technische Hochschule}
\acro{S50}[$S_{50}$]{Intelligibility Function Gradient at the \ac{SRT}}
\acro{SA}{Spectral Amplitude}
\acro{SAA}{Synthetic Aperture Audio}
\acro{SAP}{Speech And Audio Processing}
\acro{SAR}{Speech-to-Artifact Ratio}
\acro{SAR2}[SAR]{Speaker Alternation Rate}
\acro{SAR3}[SAR]{Synthetic Aperture \ac{RADAR}}
\acro{SAS}{Synthetic Aperture \ac{SONAR}}
\acro{SB}{Subband}
\acro{SC-PHD}{Single Cluster \acs{PHD}}
\acro{SCAF}{Single Channel Adaptive Filter}
\acro{SCB}{Stochastic Codebook}
\acro{SCNR}{Single-Channel Noise Reduction}
\acro{SCOT}{Smoothed Coherence Transform}
\acro{SCNR}{Single-Channel Noise Reduction}
\acro{SC-PHD}{Single Cluster \acs{PHD}}
\acro{SCR}{Signal-to-Competition Ratio}
\acro{SCRIBE}{Spoken Corpus of British English}
\acro{SCT}{Speech Corruption Toolkit}
\acro{SCT2}{Short Conversation Test}
\acro{SD}{Semantic Differential}
\acro{SDB}{Superdirective Beamformer}
\acro{SDD}{Spectral Decay Distributions}
\acro{SDDMSB}{\acs{SDD} with Mel-spaced frequency bands}
\acro{SDDSA}{\acs{SDD} with Mel-spaced frequency bands and selective averaging}
\acro{SDDSA-G}{\acs{SDDSA} with Gerkmann noise estimator}
\acro{SDDSA-H}{\acs{SDDSA} with Hendriks noise estimator}
\acro{SDR}{Software Defined Radio}
\acro{SDR2}{Speech}
\acro{SDRAM}{Synchronous Dynamic Random Access Memory}
\acro{SDT}{Speech Description Taxonomy}
\acro{SEDF}{Subband \ac{EDF}}
\acro{SELD}{Sound Event Localization and Detection}
\acro{SEMG}{Surface Electromyography}
\acro{SFDR}{Spurious Free Dynamic Range}
\acro{SFM}{Single Feature with Mapping}
\acro{SFT}{Spherical Fourier Transform}
\acro{SH}{Spherical Harmonic}
\acro{SHD}{Spectral Harmonic Decomposition}
\acro{SHD2}[SHD]{Spherical Harmonic Domain}
\acro{SIE}{System Identification Error}
\acro{SII}{Speech Intelligibility Index}
\acro{SIImod}{Speech Intelligibility Index in the modulation domain}
\acro{SIM}{Subscriber Identity Module}
\acro{SIMO}{Single-Input-Multiple-Output}
\acro{SINAD}{Signal-to-Noise and Distortion Ratio}
\acro{SIP}{Session Initiation Protocol}
\acro{SIR}{Signal-to-Interference Ratio}
\acro{SIR2}[SIR]{Sequential Importance Resampling}
\acro{SIREAC}{Simulation of REal Acoustics\acroextra{ Software Tool}}
\acro{SIS}{Sequential Importance Sampling}
\acro{SL}{Speech Level}
\acro{SLAM}{Simultaneous Localization and Mapping}
\acro{SLLN}{Strong Law of Large Numbers}
\acro{SLM}{Sound Level Meter}
\acro{SMA}{Spherical Microphone Array}
\acro{SMARD}{Single- and Multichannel Audio Recordings Database}
\acro{SMERSH}{Spatiotemporal Averaging Method for Enhancement of Reverberant Speech}
\acro{SMIR}{Spherical Microphone array Impulse Response}
\acro{SMPTE}{Society of Motion Picture and Television Engineers}
\acro{SMS}{Short Message Service}
\acro{SNR}{Signal-to-Noise Ratio}
\acro{SNR2}[SNR]{Speech-to-Noise Ratio}
\acro{SNT}{Subspace Noise Tracking\acroextra{ algorithm}}
\acro{SOBM}{STOI-optimal Binary Mask}
\acro{SONAR}{SOund Navigation And Ranging}
\acro{SOLA}{Synchronous Overlap Add\acroextra{. A method of scaling a signal in time and pitch independently.}}
\acro{SPC}{Specificity}
\acro{SPEECON}{Speech Databases for Consumer Devices}
\acro{SPHERE}{NIST SPeech Header REsources\acroextra{ software with embedded Shorten Compression}}
\acro{SPIN}{Speech Perception In Noise}
\acro{SPL}{Sound Pressure Level}
\acro{SPP}{Speech Presence Probability}
\acro{SPQA}{Speech Quality Assurance Package}
\acro{SQNR}{Signal-to-Quantization Noise Ratio}
\acro{SR}{Sparse Representation}
\acro{SR2}[SR]{Spectral Rotation}
\acro{SRA}{Statistical Room Acoustics}
\acro{SRI}{SRI International\acroextra{. Formerly Standford Research Institute}}
\acro{SRMR}{Speech-to-Reverberation Modulation Energy Ratio}
\acro{SRP}{Steered Response Power}
\acro{SRP-PHAT}{Steered Response Power with Phase Transform}
\acro{SRP-TDE}{Steered Response Power with Time Delay Estimation}
\acro{SRR}{Signal-to-Reverberation Ratio}
\acro{SRT}{Speech Reception Threshold\acroextra{ (also known as Speech Recognition Threshold)}}
\acro{SS}{Spectral Subtraction}
\acro{SSB}{Single Side-Band}
\acro{SSI}{Synthetic Sentence Identification}
\acro{SSL}{Sound Source Localization}
\acro{SSN2}[SSN]{Simultaneous Switching Noise}
\acro{SSN}{Speech-Shaped Noise}
\acro{SSNR}{Segmental \acs{SNR}}
\acro{SSOBM}{Stochastic \acs{SOBM}}
\acro{SSRR}{Segmental Signal-to-Reverberation Ratio}
\acro{SSW}{Staggered Spondaic Word}
\acro{STFT}{Short Time Fourier Transform}
\acro{STI}{Speech Transmission Index}
\acro{STIPA}{Speech Transmission Index for Public Address Systems}
\acro{STITEL}{Speech Transmission Index for Telecommunication Systems}
\acro{STMI}{Spectro-Temporal Modulation Index}
\acro{STNR}{\acs{NIST}'s Speech-to-Noise Ratio\acroextra{ Estimation Algorithm}}
\acro{STOI}{Short-Time Objective Intelligibility\acroextra{ measure}}
\acro{STQ}{Speech Processing, Transmission and Quality Aspects}
\acro{STSA}{Short Time Spectral Analysis}
\acro{STSA1}[STSA]{Short Time Spectral Amplitude}
\acro{SUS}{Semantically Unpredictable Sentences}
\acro{SVD}{Singular Value Decomposition}
\acro{SVM}{Support Vector Machine}
\acro{SWSOBM}{Stochastic \acs{WSOBM}}
\acro{T20}[$T_\textrm{20}$]{Reverberation Time\acroextra{ to decay by $20$ dB}}
\acro{T30}[$T_\textrm{30}$]{Reverberation Time\acroextra{ to decay by $30$ dB}}
\acro{T60}[$T_\textrm{60}$]{Reverberation Time\acroextra{ to decay by $60$ dB}}
\acro{TBM}{Target Binary Mask}
\acro{TDE}{Time Delay Estimation}
\acro{TDHS}{Time Domain Harmonic Scaling\acroextra{. A method of scaling a signal in time and pitch independently.}}
\acrodefplural{TDOA}[TDOAs]{Time-Differences-of-Arrival}
\acro{TDOA}{Time-Difference-of-Arrival}
\acrodefplural{TDOA}[TDOAs]{Time-Differences-of-Arrival}
\acro{TDT}{Tone Decay Test}
\acro{TF}{Time-Frequency}
\acro{TFS}{Temporal Fine Structure}
\acro{TFGM}{Time-Frequency Gain Modification\acroextra{. An approach to signal enhancement in which a signal is multiplied by a gain function in the time-frequency domain.}}
\acro{THD}{Total Harmonic Distortion}
\acro{TI}{Texas Instruments, Inc.}
\acro{TIMIT}{\acs{TI}-\acs{MIT} speech corpus}
\acro{TIPHON}{Telecommunication and Internet Protocol Harmonization Over Networks}
\acro{TLS}{Total Least-Squares}
\acro{TN}{True Negative}
\acro{TNR}{True Negative Rate}
\acro{TOA}{Time-of-Arrival}
\acrodefplural{TOA}[TOAs]{Times-of-Arrival}
\acro{TOF}{Time-of-Flight}
\acro{TOSQA}{Telekom Objective Speech Quality Assessmentt}
\acro{TP}{True Positive}
\acro{TP2}[TP]{Trivial Pursuit}
\acro{TPCC}{Trivial Pursuit with Clipping Constraints}
\acro{TPR}{True Positive Rate}
\acro{TSE}{Taylor Series Expansion}
\acro{TTD}{Theoretical Time Delay}
\acro{TVAR}{Time-varying Autoregression}
\acro{UAV}{Unmanned Aerial Vehicle}
\acro{UDP}{User Datagram Protocol}
\acro{UFRJ}{Federal University of Rio de Janeiro}
\acro{UHF}{Ultra High Frequency}
\acro{UKF}{Unscented Kalman Filter}
\acro{ULA}{Uniform Linear Array}
\acro{ULF}{Ultra Low Frequency}
\acro{UMTS}{Universal Mobile Telecommunications Service}
\acro{US}{United States}
\acro{UTBM}{Universal Target Binary Mask}
\acro{VAD}{Voice Activity Detector}
\acro{VBR}{Variable Bit-Rate}
\acro{VCV}{Vowel-Consonant-Vowel}
\acro{VRD}{Variance of Decay-rates}
\acro{VGC}{Voice Grade Channel}
\acro{vMF}{von Mises-Fisher}
\acro{VoIP}{Voice Over Internet Protocol}
\acro{VRT}{Vlaamse Radio- en Televisieomroeporganisatie\acroextra{. (Flemish Radio and Television Broadcasting Organization)}}
\acro{VSELP}{Vector Sum-excited Linear Prediction}
\acro{VST}{Virtual Studio Technology\acroextra{. An interface standard developed by Steinberg for adding plugins to an audio editor.}}
\acro{WADA}{Waveform Amplitude Distribution Analysis}
\acro{WASN}{Wireless Acoustic Sensor Network}
\acrodefplural{WASN}[WASNs]{Wireless Acoustic Sensor Networks}
\acro{WASPAA}{Workshop on Applications of Signal Processing to Audio and Acoustics}
\acro{WAV}{Waveform Audio File Format}
\acro{WAVE}{Waveform Audio File Format}
\acro{WB}{Wideband}
\acro{WER}{Word Error Rate}
\acro{WGN}{White Gaussian Noise}
\acro{WLAN}{Wireless \acs{LAN}}
\acro{WLLN}{Weak Law of Large Numbers}
\acro{WM}{Working Memory}
\acro{WMA}{Windows Media Audio}
\acro{WNG}{White Noise Gain}
\acro{WSS}{Weighted Spectral Slope}
\acro{WSOBM}{Weighted \acs{SOBM}}
\acro{WSTOI}{Weighted \acs{STOI}}
\acro{ZOS}{Zero-Order Statistics}

%% file: sections/1_introduction.tex
\begin{abstract}
    \ac{SRP} is a widely used method for the task of sound source localization using microphone arrays, showing satisfactory localization performance on many practical scenarios. However, its performance is diminished under highly reverberant environments. Although \acfp{DNN} have been previously proposed to overcome this limitation, most are trained for a specific number of microphones with fixed spatial coordinates. This restricts their practical application on scenarios frequently observed in wireless acoustic sensor networks, where each application has an ad-hoc microphone topology. We propose Neural-SRP, a \ac{DNN} which combines the flexibility of \ac{SRP} with the performance gains of \acp{DNN}. We train our network using simulated data and transfer learning, and evaluate our approach on recorded and simulated data. Results verify that Neural-SRP's localization performance significantly outperforms the baselines.
\end{abstract}

\noindent\textbf{Index Terms}: \acf{SSL}, \acf{DNN}, \acf{SRP}, \acf{DMA}

\section{Introduction}

\begin{figure}[h]
    \centering
        \includegraphics[scale=0.6]{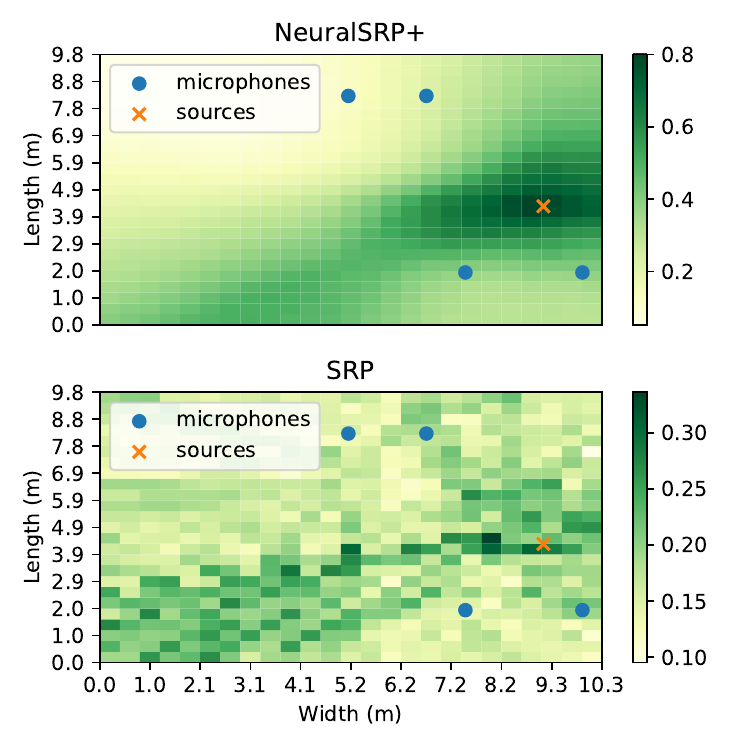}
    \caption{Neural-SRP and classical SRP output for real recorded signals.}
    \label{fig:outputs}
\end{figure}

This paper focuses on the task of \acf{SSL} using signals recorded by a \acf{DMA}. The goal of \ac{SSL} is to estimate the position of a sound source of interest, such as a human talker, in a room. Applications of \ac{SSL} include speech enhancement~\cite{brandstein2001microphone}, diarization \cite{Zheng2021}, and robot orientation \cite{Evers2020a}. In turn, a \ac{DMA} is a network of connected microphones which can be made using custom devices consumer devices such as voice assistants, cell phones, and laptops equipped with one or more microphones. When these devices are wirelessly connected, a \ac{DMA} is commonly referred to as a \acf{WASN} \cite{Bertrand2011}.

A classical approach for \ac{SSL} is to apply the \acf{SRP} method, which operates by assigning a likelihood to each location in a grid of candidate source positions. \ac{SRP} can function on \acp{DMA} of any geometry and number of microphones, and has been shown to be robust in moderately reverberant environments \cite{Zhang2008a}. However, its performance is degraded under highly reverberant environments, as its sound propagation model does not model reflections caused by walls and objects within the room.

\acf{DNN} methods have been proposed to overcome this limitation \cite{Grumiaux2021, Chakrabarty2017, Adavanne2018} by using reverberant data for training. However, as most methods were developed for centralized microphone arrays such as a single voice assistant, their application is unsuitable for \acp{WASN} where the position and number of microphones may be unknown in advance or dynamically change. A change in number may be caused by the failure of a device, or its position may be moved by a user. 

In this work we propose Neural-SRP, a method which combines the advantages of the classical \ac{SRP} method and \acp{DNN}. As in the classical \ac{SRP}, this network is able to function on unseen \ac{WASN} topologies, producing a likelihood grid as its output that is illustrated in the top panel of \autoref{fig:outputs}. As \autoref{fig:outputs} shows, the maps produced by Neural-SRP are much smoother than classical SRP, resulting in an increased localization performance. This is achieved by training the network on simulated and recorded reverberant data. Another advantage of Neural-SRP is that it does not require calibrated microphone gains, allowing a \ac{DMA} of heterogeneous devices to be used.

This paper continues as follows. In \autoref{sec:statement} the signal model and scope of this paper is formulated. Relevant background is provided in \autoref{sec:literature}, first by describing the \ac{SRP} method followed by a description of relevant work in deep learning for \ac{SSL}. The proposed method is described in \autoref{sec:neural_srp}, followed by its experimental validation in \autoref{sec:experiments}. Finally, results are presented in \autoref{sec:results} and \autoref{sec:conclusion} concludes this work.

\section{Problem statement} \label{sec:statement}

Our goal is to estimate the position of a static sound source located at an unknown position $\pmb{p}_s = [p_s^x \, p_s^y \, p_s^z]^T$ within a reverberant room of known dimensions $\pmb{d} = [d^{x} \, d^{y} \, d^{z}]^T$. The source emits a speech signal $s(t)$ at the discrete time index $t$. Besides the source, $M$ microphones are present in the room, where microphone $m$ has a known position $\pmb{p}_m = [p_m^x \, p_m^y \, p_m^z]^T$, and receives a signal $x_m(t)$ equal to 

\begin{equation} \label{eq:prop-reverb}
    x_m(t) = \sum_{r=0}^{R - 1} h_m^r s(t - r) + \epsilon_m(t),
\end{equation}
where vector $[h_m^0 \ldots h_m^{R - 1}]^T$ is the \ac{RIR} between microphone $m$ and source, and $\epsilon_m(t)$ is measurement noise. In an anechoic room, \eqref{eq:prop-reverb} can be simplified as
\begin{equation} \label{eq:received_signal_anechoic}
    x_m(t) = a_m s(t - \tau_m) + \epsilon_m(t) ,
\end{equation}
where $a_m$ and $\tau_m$ respectively represent the attenuation and delay caused by propagation. The anechoic formulation defined in \autoref{eq:received_signal_anechoic} is used by the classical \ac{SRP}, and does not account for reverberation. 


In our method and in the baselines considered in this paper, the microphone signals are sampled and processed in frames of size $L$, defined as $\pmb{x}_m(t) = [x_m(t - L + 1) \, ... \, x_m(t)]^T$ We assume synchronicity between the microphones' digital-to-analog conversion clocks, or that a sampling rate compensation algorithm \cite{Chinaev2019} can be used. Finally, we also define a metadata vector $\pmb{\phi}$ as
\begin{equation} \label{eq:phi}
    \pmb{\phi} = [\pmb{p}_1^T \, ... \, \pmb{p}_M^T \, \pmb{d}^T]^T,
\end{equation}
which specifies the microphone positions and room dimensions, which may be explored as part of the source estimation procedure.

Given the aforementioned data, the goal of our proposed method and baselines is to estimate the 2D coordinates $\hat{\pmb{p}}_s = [\hat{p}_s^x \, \hat{p}_s^y]^T$ of the source, dispensing with the need for estimating its height. A practical scenario for this assumption is work meetings, where all talkers are located at approximately the same height.

%% file: sections/2_literature_review.tex
\section{Related work} \label{sec:literature}

\subsection{Steered Response Power} \label{sec:srp}

We start by defining the \ac{SRP} between a candidate source location $\pmb{p} = [p^x \, p^y]^T$ and a pair of microphones $(i, j)$. We will omit the index $t$ for conciseness. The \textit{pairwise SRP} for a candidate location $\pmb{p}$ is defined as 
\cite{Cobos2011, DiBiase2000}
\begin{equation} \label{eq:pairwise_srp}
    \text{SRP}_{ij}(\pmb{p} \, ; \,\, \pmb{x}_i, \pmb{x}_j) = (\pmb{x}_i \star \pmb{x}_j) ( \tau_{ij}( \pmb{p})),
\end{equation}
or the cross-correlation, represented by $\star$, between frames $\pmb{x}_i$ and $\pmb{x}_j$ evaluated at the theoretical \acf{TDOA} 
\begin{equation} \label{eq:tdoa}
    \tau_{ij}(\pmb{p}) = (\lVert \pmb{p}_i - \pmb{p} \rVert - \lVert \pmb{p}_j - \pmb{p} \rVert)/c, 
\end{equation}
that is, the time propagation difference between the microphones located at $\pmb{p}_i$ and $\pmb{p}_j$ and the source $\pmb{p}$. $c$ is the speed of sound. In practice, \ac{GCC-PHAT} \cite{Knapp1976b} is commonly used instead of classical temporal cross-correlation. Finally, the global \ac{SRP} is defined as the sum between all SRP pairs: \begin{equation} \label{eq:srp}
    \text{SRP}(\pmb{p} \, ; \,\, \pmb{x}) = \sum_{i=1}^M \sum_{j=i+1}^{M} \text{SRP}_{ij}(\pmb{p} \, ; \,\, \pmb{x}_i, \pmb{x}_j)
\end{equation}

This value represents the likelihood of a source being located at a candidate point $\pmb{p}$. In other words, the source location is estimated as $ \argmax_{\pmb{p
}} \text{SRP}(\pmb{p})$. In practice, \eqref{eq:srp} is evaluated for a grid of uniformly spaced points. 



\subsection{Neural Networks for SSL}

Neural networks have been widely applied for the task of \ac{DOA} estimation using a centralized microphone array. In contrast with our application, \ac{DOA} estimation predicts the angles between the source and a reference point in the array, and range is not estimated. In \cite{Grumiaux2021}, a review of papers is provided. In \cite{Perotin2019a}, the choice of output strategy, regression or classification, is investigated. In \cite{Chakrabarty2017}, the phase of the \ac{STFT} is used as input feature to a \ac{CNN}, allowing the network to function using uncalibrated microphone gains. In \cite{He2018}, a Gaussian-like output target is introduced, which we adapt for the task of \ac{SSL}.

Works dedictated to working with a variable number of microphone positions include \cite{Furnon2021}, where an attention-based neural network capable of handling connection failures is proposed for the task of speech enhancement. However, this network is limited to a maximum number of input microphones channels. In \cite{Tzirakis2021}, a \ac{GNN} \cite{Kipf2017} is also used for speech enhancement. However, this architecure is unsuited for \ac{SSL}, as they are unable to incorporate the microphone positions in their input.

%% file: sections/3_method.tex
\section{Neural-SRP} \label{sec:neural_srp}

\begin{figure}
    \centering
        \includegraphics[scale=0.8]{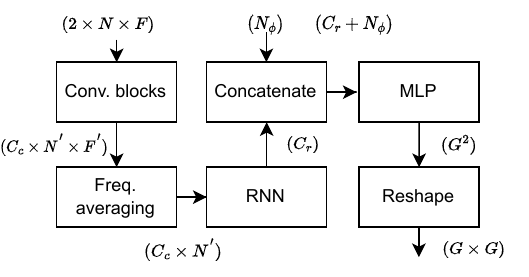}
    \caption{Architecture of the Neur al-SRP network.}
    \label{fig:neural-srp-architecture}
\end{figure}

Our proposed method works in a similar way to the classical SRP defined in \eqref{eq:srp}, that is,

\begin{equation} \label{eq:neural-srp}
    \text{NSRP}(\pmb{p} \, ; \,\, \pmb{x}) = \sum_{i=1}^M \sum_{j=i+1}^{M} \text{NSRP}_{ij}(\pmb{p} \, ; \,\, \pmb{x}_i, \pmb{x}_j),
\end{equation}

where $\text{NSRP}$ represents our neural network architecture used. We note that, in practice, $\text{NSRP}$ jointly computes a likelihood map for a square grid $\pmb{P}$ of candidate points of size $G^2$, as shown in \autoref{fig:neural-srp-architecture}.

\begin{figure}[t]
    \centering
        \includegraphics[scale=0.8]{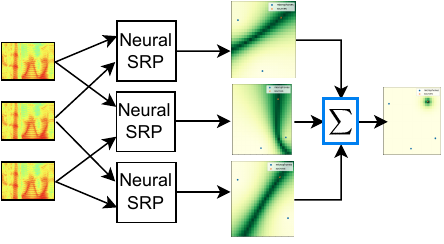}
    \caption{Example of the Neural-SRP method for three microphones. The magnitude of the STFT instead of its phase is used as input for illustrative purposes.}
    \label{fig:neural-srp-graph}
\end{figure}

An example of our approach's functioning for three microphones is shown in \autoref{fig:neural-srp-graph}. Note that however, in practice, the candidate positions $\pmb{p}$ are not provided as input, but only used to generate the training targets. The input of the network consists of a pair of signal frames $(\pmb{x}_i, \pmb{x}_j)$ and metadata $\pmb{\phi}_{ij} = [\pmb{p}_i^T \pmb{p}_j^T \pmb{d}^T]^T$, that is, a vector concatenating the coordinates of microphones $i$ and $j$ and the room dimensions $d$.

\autoref{fig:neural-srp-architecture} shows the building blocks that compose the Neural-SRP architecture, as well as their input and output shapes. Our backbone consists of a \acf{CRNN} \cite{Choi2017}, which is widely adopted for the 
task of \ac{DOA} estimation \cite{Cao2019a, Adavanne2018, Grinstein2022, Grumiaux2021a}. A feature of our architecture is its dual-input structure which allows it to incorporate the metadata vector $\pmb{\phi}_{ij}$ and allow it to function on any room and \ac{WASN} topology.

As in \cite{Chakrabarty2017}, the input of Neural-SRP is the stacked phase of the \ac{STFT} from the two microphone channels, i.e., a tensor of dimensions $(2, N, F)$, where $N$ and $F$ respectively represent the number of time and frequency bins. A \acf{CNN} is then applied to these features producing a 3D tensor of $C_c$ channels. The frequency dimension is then averaged, resulting in a 2D matrix which is input to a \ac{RNN} with \acp{GRU} \cite{Chung2014a}. We use the last output stage produced by this network, a feature vector of size $C_r$ and concatenate it to the metadata vector $\pmb{\phi}_{ij}$ of length $N_{\phi}$ to produce a \textit{metadata aware} feature. In the field of multimodal fusion \cite{Atrey2010}, the combination procedure of signals and metadata is classified as a \textit{late fusion} approach. Finally, this metadata-aware feature vector serves as input to a \ac{MLP}, which generates the output likelihood grid of size $G^2$.  

We remark that using a CNN with unitary time kernels and a uni-directional RNN makes our architecture causal, and therefore able to be employed in real-time applications.

\subsection{Training targets}

\begin{figure}[h]
    \centering
        \includegraphics[scale=0.6]{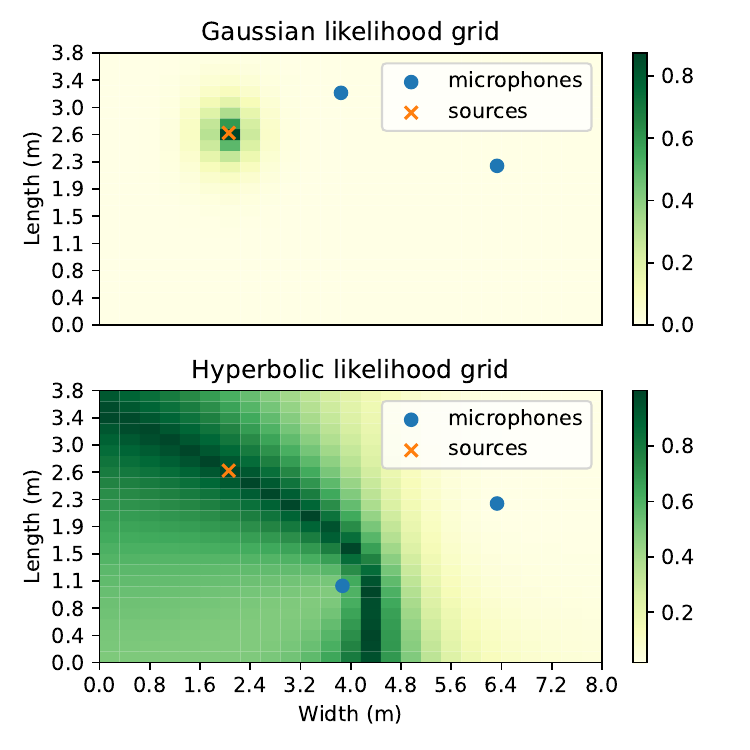}
    \caption{Example of the hyperbolic grid (below), used for training Neural-SRP, and the alternative Gaussian grid (above).}
    \label{fig:output_grids}
\end{figure}

Our network's desired output is a likelihood grid where each candidate grid cell's value is proportional to its distance to the source, that is, $\text{NSRP}(\pmb{p}) \propto |\pmb{p} - \pmb{p}_s|$. We would also like to bound the value in the interval $[0, 1]$, so it can be normalized and interpreted as a probability. In \cite{He2018}, a Pseudo-Gaussian target equal to
\begin{equation} \label{eq:gaussian_grid}
    y_g(\pmb{p}) = e^{-( \lVert \pmb{p} - \pmb{p}_s \rVert /\sigma)^2}
\end{equation}
is used, which satisfies the aforementioned properties. However, this target is unrealistic for training on a single microphone pair, as source locations share the same cross-correlation function along a hyperbola \cite{Gustafsson2003a} in anechoic scenarios. We therefore model our target output using a \textit{hyperbolic target grid}, defined for each grid cell as
\begin{equation} \label{eq:target-grid}
    y_h(\pmb{p}) = e^{-(\lVert \tau_{ij}(\pmb{p}) - \tau_{ij}(\pmb{p}_s) \rVert / \sigma)^2},
\end{equation}
where $\sigma$ controls the function's decay. $\tau_{ij}(\pmb{p})$ represents the theoretical \ac{TDOA} between microphones $i$ and $j$ from a source at $\pmb{p}$, as defined in \eqref{eq:tdoa}. A comparison between both grids can be seen in \autoref{fig:output_grids}. Note that \eqref{eq:target-grid} is maximized to a value of 1 at the source's location, and decays to 0 as the distance to the source increases. This grid assigns a high likelihood value along a hyperbola branch, hence its name. An interpretation for this target is that the network must learn to estimate a likelihood grid similar to one that would be produced by SRP in an anechoic environment. In other words, the network has to learn to jointly dereverberate and locate the source.

Finally, we define the training loss function as
\begin{equation} \label{eq:loss}
    \mathcal{L}(\pmb{Y}, \hat{\pmb{Y}}) = |\pmb{Y} -\hat{\pmb{Y}}|,
\end{equation}

the \ac{MAE} between the target grid $\pmb{Y}$ computed using \eqref{eq:target-grid} and the network output $\hat{\pmb{Y}}$. The training procedure for Neural-SRP consists of applying a gradient based method search for the network parameters or weights which minimize \eqref{eq:loss}.

\subsection{Anechoic to reverberant transfer learning} \label{sec:transfer-learning}
We found that training directly on highly reverberant data led to the network getting stuck on local optima values. We were able to overcome this by applying a two-stage training procedure. Firstly, the network is trained using anechoic data. Then, we resume training using a reverberant dataset, which allowed the network to reach substantially lower error values. This procedure is referred to as transfer learning or fine-tuning \cite{Tan2018}.

%% file: sections/4_experimentation.tex
\section{Experimentation} \label{sec:experiments}

\subsection{Datasets}
This section describes the different datasets used for training and evaluation of Neural-SRP and the baselines. To assess the dependence of Neural-SRP's performance on the number of microphones used, two testing variants of each dataset were produced, one containing 4 microphones and one containing 6 microphones. Furthermore, to train the baseline \ac{CRNN}, training and validation datasets containing 4 and 6 microphones were used. We emphasize that no combination of microphone and source positions overlap between training, validation and testing datasets. The microphone signals used had a duration of 0.5~seconds.  

\vspace{0.3cm}

The \textbf{AnechoicSim} dataset is a simulated dataset with no reverberation, used to train the first stage of the network. It consists of $10000$ training, $2500$ validation and $2500$ test samples. The source signal used are speech samples from the VCTK corpus \cite{Yamagishi2019a}. For each dataset sample, the room's width, length and height are randomly uniformly chosen from the respective intervals of $[3, 10]$, $[3, 10]$ and $[2, 4]$ meters. The source and microphone positions were uniformly sampled within the room's dimensions, with the restriction of each device pair being at least $0.5$~m apart.

\vspace{0.3cm}

The \textbf{ReverbSim} is a dataset generated using the Image Source method \cite{Allen1979a} and speech samples from the VCTK corpus, used to train the network at a second stage. Additionally to using the same random room dimension, microphone and source positions ranges of AnechoicSim, the reverberation of each room is set using the \textit{reflectivity-biased} procedure \cite{Foy2021}, which assigns one absorption coefficient per surface, generating more realistic rooms than creating surfaces with a shared coefficient. It consists of $10000$ training, $2500$ validation and $2500$ test samples.

\vspace{0.3cm}

The \textbf{Recorded} dataset \cite{Guan2021} contains real recordings from a single room with a high reverberation time of $800$~ms containing 40 microphones and 4 loudspeaker positions. The loudspeakers emit sounds from the LibriSpeech \cite{Panayotov2015b} corpus. This dataset was used for evaluating the proposed model and baselines, as well as for fine-tuning the neural models. It consists of 250 training samples and 2500 testing samples.

\subsection{Methods and baselines}

We evaluate our approach against the classical SRP method using a similar approach to  \cite{Cobos2011}. Furthermore, we experiment training Neural-SRP using two or three stages. The two-stage approach, which we will refer as \textit{NeuralSRP}, consists of training the network using simulated data as defined in \autoref{sec:transfer-learning}, firstly by training the network on the AnechoicSim dataset, followed by training it on the ReverbSim dataset. The three-stage approach, denoted \textit{NeuralSRP+}, is further training the \textit{NeuralSRP} model using a small subset of the Recorded dataset. Using recorded data to complement synthetic training was used in other works such as \cite{He2018}, as simulated data may not completely match real scenarios.

To compare Neural-SRP against a trained baseline, we propose the CRNN4 and CRNN6 models, which share a similar architecture to and number of parameters to NeuralSRP, but differ in the input and output. CRNN4 and CRNN6 jointly process respectively 4 and 6 microphone signals, and therefore have to be trained specifically for each case. The architecture used is therefore the same as shown in \autoref{fig:neural-srp-architecture}, but the input to the Convolutional blocks is $(4 \times N \times F)$ for CRNN4, and $(6 \times N \times F)$ for CRNN6. Furthermore, a single output grid is produced instead of summing grids from all pairs. For that reason, we train the aforementioned models using the Pseudo-Gaussian target as described in \eqref{eq:gaussian_grid}.
\subsection{Experiment details}
A public repository containing all methods is provided on Github, as well as a demonstration website for data access and reproduction \footnote{\url{https://github.com/egrinstein/gnn_ssl}}. Pytorch \cite{Paszke2019a} was used as the main deep learning library, along with Pytorch Lightning \cite{Falcon2019a} for abstracting common training routines. Pyroomacoustics \cite{Scheibler2018a} was used for generating the AnechoicSim and ReverbSim datasets.

We train the networks for a maximum of 50 epochs with early stopping if the validation loss stops increasing after 3 epochs. We employ a learning rate of 0.0005 using the Adam optimizer \cite{Kingma2017}. We use a batch size of 16. These parameters were chosen empirically. All grids used are of dimensions $25 \times 25$. The neural networks contain four convolutional layers using $1\times 2$ kernels and filter sizes and a MLP containing 3 layers, each of output size 625. The total number of parameters for the neural network methods is around one million. We use a ReLU activation function for all layers except for the output, which uses no activation. The source estimation procedure in the baselines and proposed methods consist of picking the location of the highest value, and \ac{GCC-PHAT} is used to compute the cross-correlation within the SRP baseline.

%% file: sections/5_results_and_discussion_and_conclusion.tex
\section{Results and discussion} \label{sec:results}

We employ the error $\lVert \hat{\pmb{p}} - \pmb{p} \rVert$, i.e. the distance in meters between the predicted and actual source positions, as the main metric of comparison. As a global comparison metric, the average error for all samples is computed. A table comparing our proposed methods and baselines is shown in \autoref{tab:results}. 

The highlighted results in \autoref{tab:results} show that NeuralSRP and NeuralSRP+ obtain the best results in terms of average localization error, overcoming the classical SRP method. The relative improvement between Neural and classical SRP was of 59, 54, 67, and 34\% for the ReverbSim4, Recorded4, ReverbSim6, and Recorded6 datasets respectively. Interestingly, the CRNN baselines were unable to surpass the performance of SRP. An explanation for this performance may be that the Gaussian grid is less effective for training due to its sparse nature. The performance of NeuralSRP increased when using 6 microphones in comparison to 4, as expected.

In qualitative terms, an example comparison between grids generated by classical and Neural-SRP can be seen in \autoref{fig:outputs}, showing that while SRP produces a noisy grid with peaks not located at the true source position, Neural-SRP produces a smooth grid with the maximum located at the true source's position.

\begin{table}[]

    \centering
    \begin{tabular}{c|c|c|c}
        Dataset & Model & Avg. err. (m) & Std. (m) \\
        \hline
        ReverbSim4 & SRP & 1.86 & 1.46 \\
        ReverbSim4 & NeuralSRP & \textbf{1.17} & 0.87 \\
        ReverbSim4 & CRNN4 & 2.85 & 1.53\\
        \hline
        ReverbSim6 & SRP & 1.51 & 1.32 \\
        ReverbSim6 & NeuralSRP & \textbf{0.90} & 0.66 \\
        ReverbSim6 & CRNN6 & 2.49 & 1.29 \\
        \hline
        Recorded4 & SRP & 1.19 & 1.53 \\
        Recorded4 & NeuralSRP+ & \textbf{0.77} & 0.66 \\
        Recorded4 & CRNN4 & 3.78 & 1.63\\
        \hline
        Recorded6 & SRP & 0.75 & 1.04 \\
        Recorded6 & NeuralSRP+ & \textbf{0.56} & 0.46\\
        Recorded6 & CRNN6 & 2.91 & 1.77
    \end{tabular}

    \caption{Comparison of the mean and standard deviation of the localization error between multiple datasets and models.}
    \label{tab:results}
\end{table}

\section{Conclusion and future work} \label{sec:conclusion}
In this work, we presented Neural-SRP, a method for \ac{SSL} combining the advantages of the classical \ac{SRP} method and neural networks. We show that Neural-SRP is able to operate on multiple array geometries, overcoming SRP and a CRNN baseline by a minimum margin of 34\% on multiple challenging scenarios. Future work directions include testing Neural-SRP for localizing multiple sources and performing in non-shoebox rooms.